\documentclass[12pt]{article}
\usepackage{amsthm,amsmath,amssymb}
\usepackage{mathrsfs}
\allowdisplaybreaks[3]
\usepackage{geometry}
\usepackage{graphicx}
\usepackage{hyperref}
\usepackage{cancel}
\usepackage{textcomp}
\usepackage{tikz}
\usepackage{bm}
\usepackage{mathrsfs}
\usepackage{filecontents}
\usepackage{times}
\usepackage{epsfig}
\usepackage{xcolor}
\usepackage{slashed}
\usepackage{latexsym}
\usepackage{verbatim}
\usepackage{extarrows}
\usepackage{cite}
\usepackage{multirow}
\usepackage{rotating}
\usepackage{colortbl}
\usepackage{indentfirst}
\usepackage{tensor}
\definecolor{mygray}{gray}{.9}
\definecolor{intnull}{RGB}{213,229,255}
\usepackage{arydshln}
\usepackage{diagbox}

\usepackage{caption}
\usepackage{tikz}
\usetikzlibrary{arrows,shapes,chains}
\usepackage{soul}
\usepackage{graphicx, subfig}
\begin{document}
\renewcommand{\thefootnote}{\fnsymbol{footnote}}
\baselineskip=16pt
\pagenumbering{arabic}
\vspace{1.0cm}
\begin{center}
{\Large\sf  The generalized holographic $c$-function for regular AdS black holes}
\\[10pt]
\vspace{.5 cm}

{Yang Li\footnote{E-mail address: 2120190123@mail.nankai.edu.cn} and Yan-Gang Miao\footnote{Corresponding author. E-mail address: miaoyg@nankai.edu.cn}}
\vspace{3mm}

{School of Physics, Nankai University, Tianjin 300071, China}

\vspace{4.0ex}
\end{center}
\begin{center}
{\bf Abstract}
\end{center}

We use the causal horizon entropy to study the asymptotic behaviors of regular AdS black holes. In some literature, the causal horizon entropy is regarded as a generalized holographic $c$-function. In this paper, we apply this idea to the case of regular AdS black holes. We show that the causal horizon entropy decreases to zero at the center of regular AdS black holes and in particular it is stationary because its derivative with respect to the affine parameter approaches zero asymptotically. Meanwhile, the asymptotic behavior of the metric of regular AdS black holes implies that the black hole center corresponds to an IR fixed point. Therefore, we conclude that the causal horizon entropy is a valid candidate for the holographic $c$-function of these regular AdS black holes.

\section{Introduction}
\label{sec:intr}

As a mysterious gravitational object, the black hole (BH) has been one of the central topics in modern physics. Many BH models have been proposed during the past century, some of which have observational evidences~\cite{Will,Yunes,LIGO,image,X-ray}. However, these evidences provide the information about the BH exterior region at most, and so far there has been no much information known about the interior of BHs.

Meanwhile, the bizarre nature of singularity is so compelling that many researchers cannot help but suppose that whether the singularity is an artificial consequence of gravitational models or not because no physical laws maintain valid at the singularity. In fact, as early as 1968, Bardeen proposed~\cite{Bardeen} the first regular black hole (RBH) without singularity at the center. The Bardeen BH was regarded as a mathematical construction at its beginning, then in the first year of the twenty-first century Ay\'{o}n-Beato and Garc\'{i}a (ABG) found~\cite{Bardeen_ABG} its action which is the Einstein gravity coupled to nonlinear electrodynamics (NED). This success inspired the studies of RBHs, especially those coupled with different kinds of NED fields. Different types of RBHs have been constructed ever since, {\em e.g.}, the Bardeen-type BH~\cite{Bardeen_electric,Bardeen_D,Bardeen_rotate}, the Hayward BH~\cite{Hayward,Hayward_ZYF}, the ABG-type BHs~\cite{ABG,ABGSTVG}, the RBH in Einstein-Gauss-Bonnet gravity coupled with NED fields~\cite{EGBNED}, the nonminimally coupled gravitational model of Yang-Mills fields~\cite{NEYM}, and the others~\cite{MOG,BV}. In all these RBH models, their centers are a de Sitter or an anti-de Sitter core, instead of spacetime singularity. Besides, another family of RBHs related to conformal gravity has also been constructed~\cite{Bambi003,Bambi033,Bambi1711
,Pisin,1705,2102}.

These constructions of RBHs arouse our curiosity about the nature of their center called regular center. It would be interesting to know what role the regular center plays in a gravitational theory. There have already been some methods to probe the interior of a BH, especially those proposed in the frame of AdS/CFT correspondence~\cite{AdS/CFT}. One powerful and successful approach in probing the interior of BHs is the correlation function~\cite{Fidkowski,Festuccia,Hamilton1,Hamilton2,
Hartman,Frenkel,Matan,Gomez}. The function is  divergent~\cite{Hamilton2} at the singularity and is dominated~\cite{Fidkowski} by the spacelike geodesics for the large mass bulk field. However, we find this approach does not serve for our purpose of studying RBHs because the RBHs usually have complicated mathematical forms of metrics while the approach of correlation functions requires simple forms of metrics for analytic expressions. So,  the application of correlation functions on RBHs is not feasible.

Another enchanting method to probe the center of BHs is the holographic renormalization group (RG) flow~\cite{Myers1,Ghosh,Boer,YZL} and the holographic $c$-function~\cite{Gushterov,Cremonini,Paulos,
Myers3} because the central region of a BH corresponds to the IR limit of the dual field theory of AdS/CFT correspondence. We explain this in more details.
Since Wilson's original idea was proposd~\cite{Wilson1,Wilson2}, the RG flow has been an essential constitution of quantum field theory and a powerful tool to study the relations of various quantum theories in different energy scales. One quantitative description of RG flow is $c$-function. The RG flow starts at the UV fixed point and ends at the IR fixed point. Each fixed point represents a conformal field theory which is invariant under RG transformation. Based on this fact, Zamolodchikov~\cite{Zamolodchikov} proved the existence of $c$-functions in a two-dimensional field theory and set up $c$-theorem.
This $c$-function equals the central charges of the two-dimensional CFT at the UV and IR endpoints, monotonically decreases from the UV point to the IR point, and is stationary at these two endpoints.

In general, the central charge is related to the energy momentum tensor, and there exist  more than one central charge in higher dimensions,
\begin{equation}
\langle T_{\mu}{}^{\mu}\rangle \sim -aE^{(2n)}+\sum c_i I_i^{(2n)},
\label{central charge}
\end{equation}
where $a$ is $a$-type of central charges and $c_i$'s belong to $c$-type of central charges,
 \begin{equation}
E^{(2n)} \equiv \frac{1}{2^n}\epsilon_{\mu_{1} \nu_{1} \mu_{2} \nu_{2} \cdots \mu_{n} \nu_{n}} R^{\mu_{1} \nu_{1}}\wedge R^{\mu_{2}\nu_{2}} \wedge \cdots \wedge R^{\mu_{n} \nu_{n}}
\end{equation}
 is Euler density in $2n$-dimensional spacetimes, and $I^{(2n)}$ is the collect of Weyl invariants built of Weyl tensor $C_{\mu\nu\rho\sigma}$. From Eq.~(\ref{central charge}), one can find that there exist no central charges in the odd dimensional field theory. One can also expect that the proof of the $c$-theorem is more difficult in higher dimensions. Indeed, only the four-dimensional $c$-theorem has been conjectured by Cardy~\cite{Cardy} and proved later by Komargodski
and Schwimmer~\cite{Komargodski}, where only $a$-theorem exists for the $a$-type of central charges because the $c$-function for the $c$-type of central charges in four dimensions is non-monotonic. The RG flow has also been studied through holographic approaches, see {\em e.g.}, Refs.~\cite{Fukuma,Bhattacharyya,Liu,Myers4}. In particular, the $c$-function or $c$-theorem can also be proved~\cite{Casini1,Casini2,Casini3} in holographic approaches (here the notations of $c$-function and $c$-theorem are due to the historical convention). Finally, we want to mention that a  holographic $c$-function is also closely related to the holographic entanglement entropy~\cite{HongLiu} and there is an $F$-theorem in three dimensions which describes~\cite{Casini2} the RG flow in the dual field theory of AdS/CFT correspondence, despite the absence of central charges (see also Refs.~\cite{Jafferis,Giombi}).

In the present work, we use a special kind of holographic $c$-functions to study the behavior  of the RG flow of RBHs. We connect such a kind of holographic $c$-functions to the causal horizon entropy of RBHs after being inspired by Refs.~\cite{CH1,CH2,CH3}. A causal horizon is the causal boundary of an observer who moves in a spacetime, and it carries entropy. The $c$-function we adopt here can also be understood as a general measurement of quantum correlations and quantum degrees of freedom in  a bulk spacetime~\cite{Paulos,Myers1}. We shall show that no similar pathological IR behaviors (the loss of quantum correlations) to those of Schwarzschild BHs and singular black branes (see Ref.~\cite{CH3}) exist in RBH spacetimes. According to the asymptotic form of the metric of a regular AdS black hole, the center of this black hole can be interpreted as an IR fixed point in the dual field theory of AdS/CFT correspondence, and the causal horizon entropy is a candidate for the holographic $c$-function in the regular AdS black hole spacetime.

The remaining contexts are arranged as follows. In section~\ref{sec:general}, we introduce the general formalism of causal horizon entropy. In section~\ref{sec:RBH}, we apply this formalism to four types of AdS BHs, where each of them has a regular center. This section is divided into four subsections, and each subsection focuses on one type of RBHs. The first two types are solutions of Einstein's gravity coupled with different kinds of NED fields, that is, the Hayward AdS BH in subsection~\ref{sec:Hayward} and Bardeen AdS BH in subsection ~\ref{sec:Bardeen}. The third type is the RBH in nonminimal Einstein-Yang-Mills theory in subsection~\ref{sec:NEYM} and the fourth type is the RBH in Einstein-Gauss-Bonnet gravity coupled with nonlinear electrodynamics in subsection~\ref{sec:EGBNED}. We discuss the IR behavior of the Schwarzschild BH (a singular BH) as a comparison in section~\ref{sec:singular BH}. Finally, we present our conclusions and further discussions in section~\ref{sec:concl}.

\section{General formalism: Causal horizon and its entropy}
\label{sec:general}

In this section, we introduce a formalism which describes the causal horizon of an ingoing particle. This formalism which is parallel to that in Ref.~\cite{CH3} is generalized to be suitable for RBHs.

The original metric ansatz of a spherically symmetric AdS BH is
\begin{equation}
{\rm d}s^2=-f(r){\rm d}t^2+\frac{1}{f(r)}{\rm d}r^2+r^2{\rm d}\Omega_{D-2}^2.
\label{metric ansatz r}
\end{equation}
The coordinates are $(t,r,\theta_1,\ldots,\theta_{D-2})$, where $0<\theta_1,\ldots,\theta_{D-3}<\pi$ and $0<\theta_{D-2}<2\pi$. Any spherically symmetric asymptotic AdS metrics must satisfy $\mathop{\mathrm{lim}}\limits_{r \rightarrow \infty} f(r)=r^2/\ell_{\rm AdS}^2$, where $\ell_{\rm AdS}$ is the AdS radius.

In order to study the ingoing particle which starts from the AdS boundary, we perform a coordinate transformation, $r \rightarrow 1/z$, such that Eq.~(\ref{metric ansatz r}) becomes
\begin{equation}
{\rm d}s^2=\frac{1}{z^2}\left[-h(z){\rm d}t^2+\frac{1}{h(z)}{\rm d}z^2+{\rm d}\Omega_{D-2}^2\right],
\label{metric ansatz z}
\end{equation}
where $h(z) \equiv z^2f(1/z)$.
Therefore, $\mathop{\mathrm{lim}}\limits_{z \rightarrow 0} h(z)=1$ on the AdS boundary.

The merits of Eq.~(\ref{metric ansatz z}) are twofold. At first, Eq.~(\ref{metric ansatz z}) is described in the perspective of an ingoing particle starting from the AdS boundary. The AdS boundary (departure position) corresponds to $z=0$, and the BH center to $z \rightarrow \infty$  (destiny). Secondly, according to the holographic interpretation, the $z$-coordinate naturally manifests the direction of holographic RG flows, such that $z \rightarrow 0$ corresponds to the UV behavior, and $z \rightarrow \infty$ to the IR behavior. As a result, Eq.~(\ref{metric ansatz z}) is a convenient parametrization of the original spacetime geometry.

We consider a particle whose initial position is on the AdS boundary ($z=0$), and it moves towards the BH center ($z \rightarrow \infty$). Its causal horizon is the causal boundary of motion which cannot be exceeded by the particle during its propagation. The causal horizon is generated by null Killing vectors, so it consists of the null geodesics which start from the initial position of the particle on the AdS boundary. Since we are dealing with the null geodesics, we can utilize an alternative metric which is conformally related to Eq.~(\ref{metric ansatz z}),
\begin{equation}
{\rm d}\tilde{s}^2=-h(z){\rm d}t^2+\frac{1}{h(z)}{\rm d}z^2+{\rm d}\Omega_{D-2}^2.
\label{metric conformal 0}
\end{equation}
We further recast Eq.~(\ref{metric conformal 0}) into the form,
\begin{equation}
{\rm d}\tilde{s}^2=-h(z){\rm d}v^2-2{\rm d}v{\rm d}z+{\rm d}\Omega_{D-2}^2,
\label{metric ansatz conformal}
\end{equation}
where $v\equiv t-z^*$ is the ingoing Eddington-Finkelstein coordinate, and ${\rm d}z^* \equiv {\rm d}z/h(z)$.

The equation of motion of null geodesics can be obtained as follows. The tangential vector of null geodesics has vanishing inner product,
\begin{equation}
\tilde{g}_{_{AB}}\frac{{\rm d}x^A}{{\rm d}\lambda}\frac{{\rm d}x^B}{{\rm d}\lambda}=0,
\label{null eq}
\end{equation}
where $\tilde{g}_{_{AB}}$ corresponds to Eq.~(\ref{metric ansatz conformal}) and $\lambda$ represents the null affine parameter. The R.H.S. of Eq.~(\ref{null eq}) is the Lagrangian of a null particle which corresponds to those null geodesics. The symmetry of spacetime leads to the following equations,
\begin{subequations}
\begin{eqnarray}
h(z)\frac{{\rm d}v}{{\rm d}\lambda}&=&E-\frac{{\rm d}z}{{\rm d}\lambda},
\label{charge of motion 1}
\\
\frac{{\rm d}y^i}{{\rm d}\lambda}&=&-p^i,
\label{charge of motion 2}
\end{eqnarray}
\end{subequations}
where $E$ and $p^i$, $i=1,\ldots,D-2$, are the conserved charges of motion. Note that we adopt the orthonormal coordinates $y^i$'s that satisfy ${\rm d}\Omega_{D-2}^2={\rm d}{\bf y}_{D-2}^2={\rm d}y^i{\rm d}y_i$. Substituting Eqs.~(\ref{charge of motion 1}) and (\ref{charge of motion 2}) into Eq.~(\ref{null eq}), we obtain
\begin{equation}
\left(\frac{{\rm d}z}{{\rm d}\lambda}\right)^2=E^2-p^2h(z),
\label{eq of z}
\end{equation}
where $p^2=p^ip_i$. 
Since $\mathop{\mathrm{lim}}\limits_{z \rightarrow 0} h(z)=1$, it is obvious that the conserved charges of motion must satisfy the condition, $E^2-p^2 \geq 0$. Therefore, we can parametrize $E$ and $p^i$ as follows:
\begin{subequations}
\begin{eqnarray}
E&=&\alpha \cosh \eta,\\
p^i&=&\alpha \, \hat{n}^i \sinh \eta,
\label{p parametrization}
\end{eqnarray}
\label{E p parametrization}
\end{subequations}
where $\alpha >0$, $\eta \in \mathbb{R}$, and $\hat{n}^i$ is a vector with unit length in $\mathbb{R}^{D-2}$. In fact, we can further set $\alpha=1$ because the redefinition of the affine parameter $\alpha \lambda \rightarrow \lambda$ does not change the geodesic physics. From Eqs.~(\ref{charge of motion 2}) and (\ref{p parametrization}), we have
\begin{equation}
y^i=-\lambda \, \hat{n}^i \sinh \eta.
\end{equation}
Substituting Eq.~(\ref{E p parametrization}) into Eq.~(\ref{eq of z}), we derive
\begin{equation}
\left(\frac{{\rm d}z}{{\rm d}\lambda}\right)^2=1+\left[1-h(z)\right]\sinh^2 \eta.
\label{eq of z reparametrized}
\end{equation}

Here we make an argument about Eq.~(\ref{eq of z reparametrized}). For an arbitrary $h(z)$, the R.H.S. of Eq.~(\ref{eq of z reparametrized}) can admit zero points. Consider the situation of one zero point $z_0$, such that $1+\left[1-h(z_0)\right]\sinh^2 \eta=({\rm d}z/{\rm d}\lambda)^2|_{z=z_0}=0$. At $z=z_0$, the proper speed of particle becomes zero. In fact, $z_0$ is the turning point of the particle motion. Beyond this point ($z>z_0$), $1+\left[1-h(z)\right]\sinh^2 \eta$ becomes negative such that the proper speed in $z$-direction, {\em i.e.}, ${\rm d}z/{\rm d}\lambda$, becomes imaginary. According to classical mechanics, the particle motion is usually forbidden in this region. Nevertheless, there is one exception that is the classical picture of quantum tunnelling. In quantum mechanics, there is probability for a particle to tunnel through a potential barrier. In the classical picture of quantum tunnelling, the particle motion is analytically continued into the ``classically forbidden" region. The essence is to perform a Wick rotation $t \rightarrow -i\tau$, where $\tau$ is the time inside the forbidden region. Therefore, the classical picture of quantum tunnelling is an ``imaginary-time" particle (instanton) which travels inside the forbidden region. The action of the instanton has Euclidean signature and it determines the classical motion path of the instanton.

This inspires us to analytically continue Eq.~(\ref{eq of z reparametrized}) into the ``classical forbidden region". As an analogue of the classical instanton picture, we perform $\lambda \rightarrow i\lambda$ for the region of  $1+\left[1-h(z)\right]\sinh^2 \eta < 0$, such that the non-negativity of $({\rm d}z/{\rm d}\lambda)^2$ is guaranteed. In such a sense, Eq.~(\ref{eq of z reparametrized}) becomes a piecewise function, and it can be written uniformly as follows:
\begin{equation}
\left(\frac{{\rm d}z}{{\rm d}\lambda}\right)^2=\left|1+\left[1-h(z)\right]\sinh^2 \eta\right|.
\label{eq of z absolute}
\end{equation}
The situation for more than one zero point can be generalized straightforwardly. Based on Eq.~(\ref{eq of z absolute}), the null particle can reach the center of RBHs, i.e., the causal horizon can go deep into the center. This technique of analytic continuation allows us to probe the central region of RBHs. Now we can write the proper velocity,
\begin{equation}
\frac{{\rm d}z}{{\rm d}\lambda}=\sqrt{\left|1+\left[1-h(z)\right]\sinh^2 \eta \right|},
\label{eq of z 2}
\end{equation}
where we put the absolute value symbol inside the square root to make sure that the proper velocity ${\rm d}z/{\rm d}\lambda$ is real in any spacetime regions. Thus, the affine parameter $\lambda$ has an integration expression,
\begin{equation}
\lambda=\int^z_0\frac{{\rm d}z_1}{\sqrt{\left|1+\left[1-h(z_1)\right]\sinh^2 \eta \right|}}.
\label{lambda}
\end{equation}

Next we consider the geometry of causal horizons. The causal horizon is a codimension-$1$ hypersurface, so its slice at a fixed $\lambda$ is a codimension-$2$ hypersurface, which implies that we can parametrize the latter with $D-2$ coordinates, $(\eta,n^1,\ldots,n^{D-3})$, where $(n^1,\ldots,n^{D-3})$ are $D-3$ independent components in the vector $n^i$. Therefore, the induced metric of the causal horizon at a fixed $\lambda$ has the following form,
\begin{equation}
{\rm d}s_{\rm ind}^2=\frac{1}{z^2}\left[-h(z)\left(\frac{{\rm d}v}{{\rm d}\eta}\right)^2-2\left(\frac{{\rm d}v}{{\rm d}\eta}\right)\left(\frac{{\rm d}z}{{\rm d}\eta}\right)+\lambda^2\cosh^2\eta\right]{\rm d}\eta^2+\frac{1}{z^2}\lambda^2\sinh^2\eta{\rm d}\Omega_{D-3}^2,
\label{induced metric}
\end{equation}
where ${\rm d}\Omega_{D-3}^2={\rm d}n_1^2+\cdots+{\rm d}n_{D-3}^2$.
The differential volume of the causal horizon at a fixed $\lambda$ can be expressed as
\begin{eqnarray}
{\rm d}V_{\rm ind}&=&H(\lambda,\eta){\rm d}V_{\mathbb{H}^{D-2}},
\label{induced volume}
\end{eqnarray}
with
\begin{eqnarray}
{\rm d}V_{\mathbb{H}^{D-2}}&=&\sinh^2\eta{\rm d}\eta{\rm d}\Omega_{D-3}^2
\label{H^3 volume}
\end{eqnarray}
and
\begin{equation}
H(\lambda,\eta)=\frac{\lambda^{D-3}}{z^{D-2}}\sqrt{\left|-h(z)\left(\frac{{\rm d}v}{{\rm d}\eta}\right)^2-2\left(\frac{{\rm d}v}{{\rm d}\eta}\right)\left(\frac{{\rm d}z}{{\rm d}\eta}\right)+\lambda^2\cosh^2\eta\right|},
\label{H function}
\end{equation}
where ${\rm d}V_{\mathbb{H}^{D-2}}$ is the differential volume of a $(D-2)$-dimensional unit hyperboloid $\mathbb{H}^{D-2}$. In Einstein's gravity, the entropy is evaluated by the Bekenstein-Hawking formula, $S=A/4$. Thus, the causal horizon entropy can be expressed as  $c(\lambda,\eta)=H(\lambda,\eta)/4$.

In the final part of this section, we list the formulas of ${\rm d}v/{\rm d}\eta$ and ${\rm d}z/{\rm d}\eta$. As shown in Eqs.~(\ref{charge of motion 1}) and (\ref{eq of z 2}), we have the expressions of ${\rm d}v/{\rm d}\lambda$ and ${\rm d}z/{\rm d}\lambda$. The expressions of ${\rm d}v/{\rm d}\eta$ and ${\rm d}z/{\rm d}\eta$ can be derived as follows. Starting from Eq.~(\ref{eq of z 2}), we obtain
\begin{equation}
\frac{{\rm d}}{{\rm d}\eta}\frac{{\rm d}z}{{\rm d}\lambda}=\frac{{\frak s} \left[1-h(z)\right]\cosh\eta}{\sqrt{\left|1+\left[1-h(z)\right]\sinh^2 \eta \right|}},
\label{dz d eta 1}
\end{equation}
where the factor ${\frak s}$ is defined as ${\frak s} \equiv {\rm sign}[1+\left(1-h(z)\right)\sinh^2 \eta ]$, and then derive from Eq.~(\ref{dz d eta 1}),
\begin{equation}
\frac{{\rm d}z}{{\rm d}\eta}=\int_0^{\lambda}{\rm d}\lambda_1\frac{{\frak s} \left[1-h(z)\right]\cosh\eta}{\sqrt{\left|1+\left[1-h(z)\right]\sinh^2 \eta \right|}}.
\label{dz d eta 2}
\end{equation}
Moreover, substituting Eq.~(\ref{eq of z 2}) into Eq.~(\ref{dz d eta 2}), we have
\begin{equation}
\frac{{\rm d}z}{{\rm d}\eta}=\int_0^{z}{\rm d}z_1\frac{{\frak s} \left[1-h(z_1)\right]\cosh\eta}{\left|1+\left[1-h(z_1)\right]\sinh^2 \eta \right|},
\label{dz d eta 3}
\end{equation}
from which we deduce 
\begin{equation}
\frac{{\rm d}v}{{\rm d}\eta}=\int_0^{z}{\rm d}z_1\left[-\frac{1}{h(z_1)} \frac{{\frak s}  \left[1-h(z_1)\right]\cosh\eta}{\left|1+\left[1-h(z_1)\right]\sinh^2 \eta \right|}\right].
\label{dv d eta}
\end{equation}
Eqs.~(\ref{dz d eta 3}) and (\ref{dv d eta}) are the formulas that we shall use in the next section to calculate the causal horizon entropy through numerical method because the analytic results of the above integrations are unavailable for RBHs.

It is necessary to point out that we shall deal with $z$ as a hidden parameter in the numerical calculation. Since both $\lambda$ and $H(\lambda,\eta)$ are functions of $z$, {\em viz.}, $\lambda = \lambda(z)$ and $H(\lambda,\eta)=H(\lambda(z),\eta)$, we can compute $\lambda(z)$ and $H(\lambda(z),\eta)$ by dealing with the former as the variable corresponding to the horizontal axis and the latter as the variable corresponding to the vertical axis. In this way, we can plot the graph\footnote{Because of $c(\lambda,\eta)=H(\lambda,\eta)/4$, we regard $H(\lambda,\eta)$ as causal horizon entropy in the following contexts.} of $H(\lambda,\eta)$ with respect to $\lambda$.

\section{Four kinds of regular AdS black holes}
\label{sec:RBH}

In this section, we apply the formalism given in section~\ref{sec:general} to four types of RBHs. We show that each of their regular centers corresponds to an IR fixed point, and that the causal horizon entropy manifests the IR fixed point and thus it is a candidate for holographic $c$-functions. 
Among these four types of black holes, two of them are four dimensional and the other two are five dimensional. So our discussions include both even-dimensional and odd-dimensional cases.

Our central concern is about the asymptotic behaviors of causal horizon entropy, which reflects the UV and IR behaviors of the RBH spacetimes. Therefore, in the following contexts, we only discuss the asymptotic behaviors of causal horizon entropy, {\em i.e.}, the behaviors at $z \ll 1$ and $z \gg 1$.

\subsection{4D regular Hayward AdS black hole}
\label{sec:Hayward}

This subsection serves for the 4D Hayward AdS BH~\cite{Hayward}. The action of the Hayward BH has the following form,
\begin{equation}
I_1=\frac{1}{16\pi}\int {\rm d}x^4\sqrt{-g}\left(R-2\Lambda-4{\cal L}_1(F)\right),
\label{action Hayward 1}
\end{equation}
where the cosmological constant $\Lambda=-{3}/{\ell_{\rm AdS}^2}$, and ${\cal L}_1(F)$ is the Lagrangian of the nonlinear electrodynamics,
\begin{equation}
{\cal L}_1(F)=\frac{12(\sigma F)^{3/2}}{\sigma\left[1+(\sigma F)^{3/2}\right]^2},
\label{action Hayward 2}
\end{equation}
where $\sigma$ is the parameter of nonlinear electrodynamics and $F\equiv {F^{\mu\nu}F_{\mu\nu}}/{4}$ is gauge invariant. In this model, there is only one nonvanishing component of potential, $A=Q_m\cos\theta{\rm d}\phi$, where $Q_m$ is magnetic charge which can be expressed by parameter $q$, $Q_m={q^2}/{\sqrt{2\sigma}}$. Eq.~(\ref{action Hayward 1}) together with Eq.~(\ref{action Hayward 2}) gives a static  and spherically symmetric  solution whose lapse function reads
\begin{equation}
f_1(r)=1-\frac{2q^3r^2}{\sigma\left(r^3+q^3\right)}+\frac{r^2}{\ell_{\rm AdS}^2},
\label{lapse Hayward}
\end{equation}
which behaves as
\begin{equation}
f_1(r) \sim 1+\left[\frac{1}{\ell_{\rm AdS}^2}-\frac{2}{\sigma}\right]r^2, \qquad r \rightarrow 0.
\end{equation}
This lapse function indicates that the center of Hayward BHs is a dS or an AdS core
because the asymptotic behavior is sensitive to the value of $\sigma$, either asymptotic to  dS or to AdS, depending on the values of the parameters in the lapse function. Specifically, when $\sigma$ is positive and small such that $[1/\ell_{\rm AdS}^2-2/\sigma]$ is negative, the core is of dS; when $\sigma$ is large such that $[1/\ell_{\rm AdS}^2-2/\sigma]$ is positive, the core is of AdS.

\begin{figure}[h!]
	\centering
    \includegraphics[width=0.47\textwidth]{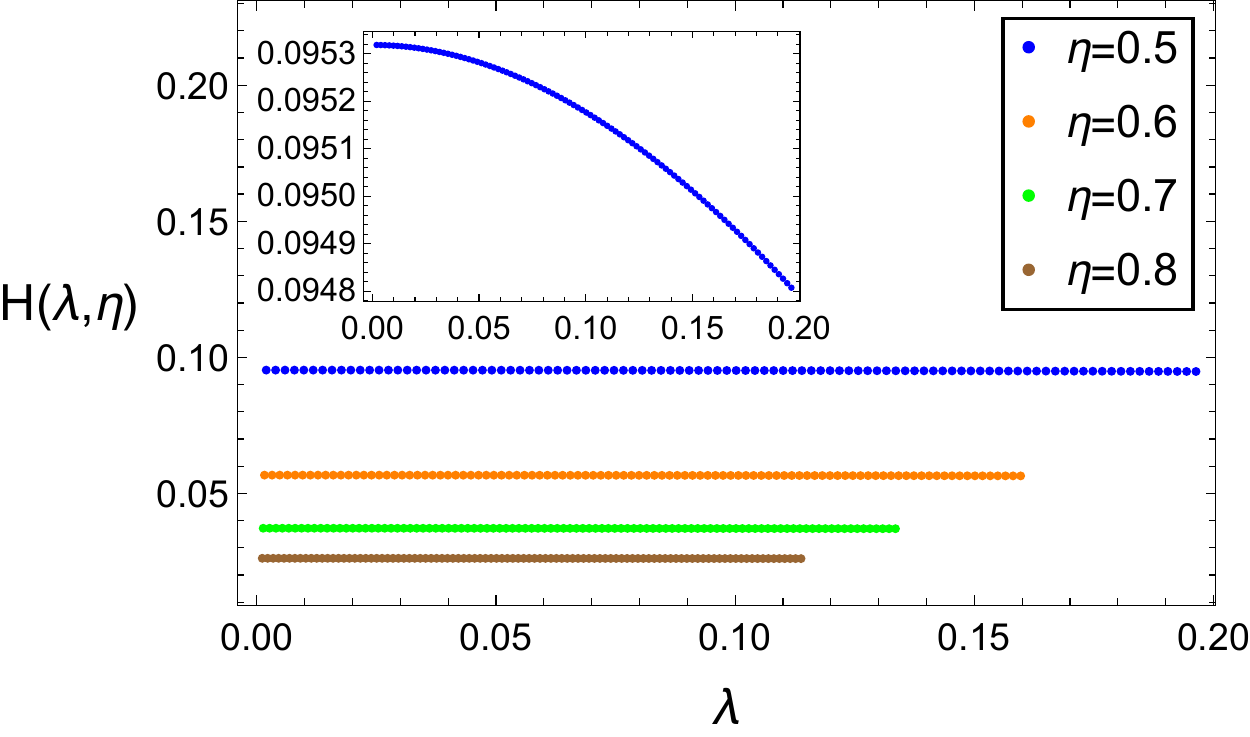}
   \includegraphics[width=0.48\textwidth]{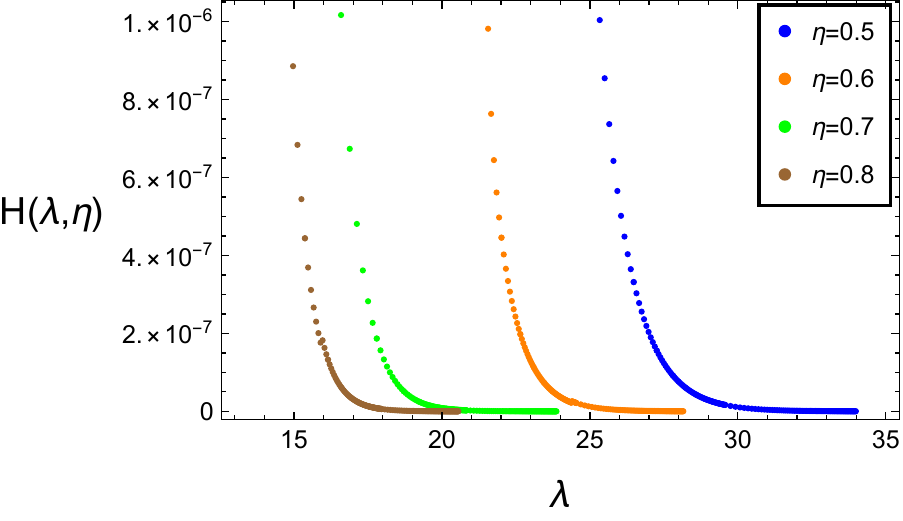}
	\caption{The left panel shows the causal horizon entropy for $z\ll 1$ (UV behavior), while the right panel for $z\gg 1$ (IR behavior). The horizontal axes of both panels represent the affine parameter $\lambda$. We have set the parameters,  $q=0.1$, $\sigma=0.1$, and $\ell_{\rm AdS}=0.1$, for the convenience of numerical calculations and plotting.}
\label{fig:c-Hayward}
\end{figure}

With the explicit expression of the lapse function Eq.~(\ref{lapse Hayward}), we are able to calculate the causal horizon entropy through the numerical method. For the Hayward AdS BH, the causal horizon entropy is the Bekenstein-Hawking entropy. Therefore, the causal horizon  entropy is $H(\lambda,\eta)$. We plot the causal horizon entropy with respect to the affine parameter for the 4D Hayward AdS BH in Fig.~\ref{fig:c-Hayward} by considering Eq.~(\ref{lapse Hayward}).

In Fig.~\ref{fig:c-Hayward}, the causal horizon entropy has vanishing derivatives in both UV and IR regions. In such a sense, we say that it is stationary in both the UV and IR limits. We conclude that the Hayward AdS spacetime possesses a UV fixed point on the AdS boundary and an IR fixed point at its regular center, and that the causal horizon entropy is a valid candidate for the holographic $c$-function of the Hayward AdS black hole.

\subsection{5D regular Bardeen AdS black hole}
\label{sec:Bardeen}

This subsection serves for the 5D Bardeen AdS BH~\cite{Bardeen_D}, which is a solution of the following action,
\begin{equation}
I_2=\frac{1}{16\pi}\int {\rm d}x^5\sqrt{-g}\left(R-2\Lambda-4{\cal L}_2(F)\right),
\label{action Bardeen 1}
\end{equation}
where the Lagrangian takes the form,
\begin{equation}
{\cal L}_2(F)=\frac{3}{4se^2}\left[\frac{\sqrt{2e^2 F}}{1+\sqrt{2e^2 F}}\right]^{7/3},
\label{action Bardeen 2}
\end{equation}
the cosmological constant $\Lambda=-{6}/{\ell_{\rm AdS}^2}$ and $4s \equiv {e^2}/{m^2}$ with charge $e$ and mass $m$. Note that $F$ has the same definition as that in the above subsection. The action Eq.~(\ref{action Bardeen 1}) together with Eq.~(\ref{action Bardeen 2}) gives a static and spherically symmetric solution, where the  lapse function reads\footnote{In Ref.~\cite{Bardeen_D} the Bardeen dS solution was given. Here we write its AdS counterpart by changing the sign of the cosmological constant.}
\begin{equation}
f_2(r)=1-\frac{m r^2}{\left(r^3+e^3\right)^{4/3}}+\frac{r^2}{\ell_{\rm AdS}^2},
\label{lapse Bardeen}
\end{equation}
and its asymptotic form is
\begin{equation}
f_2(r) \sim 1+\left[\frac{1}{\ell_{\rm AdS}^2}-\frac{m}{e^{4}}\right]r^2, \qquad r \rightarrow 0.
\end{equation}

\begin{figure}[h!]
	\centering
    \includegraphics[width=0.47\textwidth]{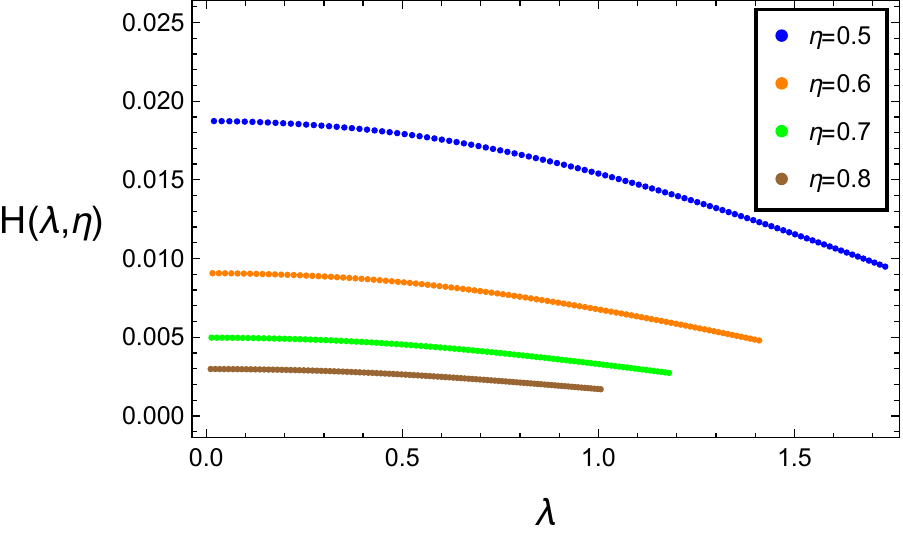}
\includegraphics[width=0.48\textwidth]{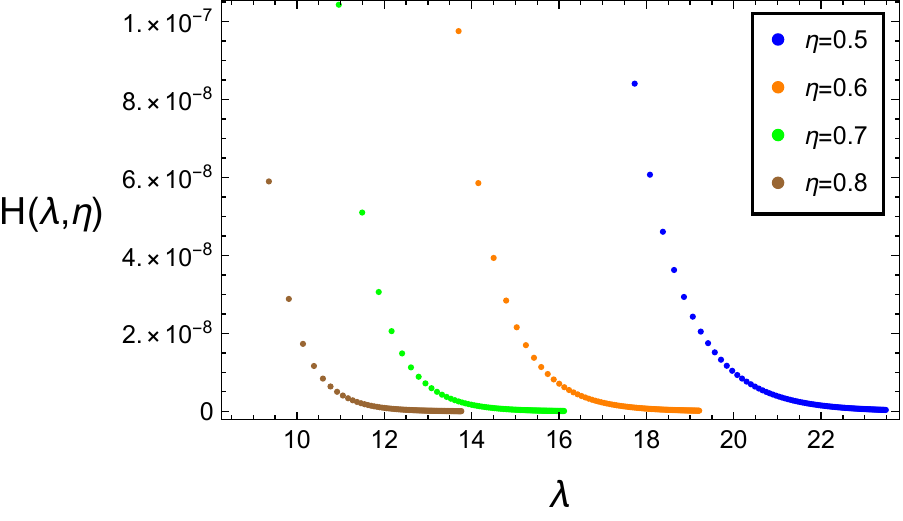}
	\caption{The left panel shows the causal horizon entropy for $z\ll 1$ (UV behavior), while the right panel for $z\gg 1$ (IR behavior). The horizontal axes of both panels are the affine parameter $\lambda$. We have set the parameters, $m=0.1$, $e=0.5$, and $\ell_{\rm AdS}=0.1$,  for the convenience of numerical calculations and plotting.}

\label{fig:c-Bardeen}
\end{figure}

Now we show the UV and IR behaviors through the numerical calculation. For the Bardeen AdS BH, the causal horizon entropy is the Bekenstein-Hawking entropy. We plot $H(\lambda,\eta)$ with respect to $\lambda$ for $z \ll 1$ and $z\gg 1$ in Fig.~\ref{fig:c-Bardeen}.

Fig.~\ref{fig:c-Bardeen} shows that the causal horizon entropy is stationary in the two asymptotic regions, {\em viz.},  $H(\lambda,\eta)$ converges smoothly to certain values. Therefore, we conclude that the causal horizon entropy is also a valid candidate for the holographic $c$-function in the case of 5D regular Bardeen AdS BHs.

The two examples in subsections~\ref{sec:Hayward} and \ref{sec:Bardeen} belong to the Einstein theory. They manifest the validity of the causal horizon entropy as a candidate for the holographic $c$-function. We shall show that the validity holds in other gravitational theories in the next two subsections.

\subsection{4D regular black hole in nonminimal Einstein-Yang-Mills theory}
\label{sec:NEYM}

This theory including nonminimally coupled Yang-Mills fields is governed~\cite{NEYM} by the following action,
\begin{equation}
I_3=\frac{1}{16\pi}\int {\rm d}x^4 \sqrt{-g}\left(R-2\Lambda+4\pi F^{(a)}_{\mu\nu}F^{(a)\mu\nu}+4\pi{\cal R}^{\mu\nu\rho\sigma}F^{(a)}_{\mu\nu}F^{(a)}_{\rho\sigma}\right),
\label{action NEYM}
\end{equation}
where the Greek indices denote the spacetime components and superscripts $(a)$ the components of gauge symmetry, $\Lambda=-{3}/{\ell_{\rm AdS}^2}$, and ${\cal R}^{\mu\nu\rho\sigma}$ is defined by
\begin{equation}
{\cal R}^{\mu\nu\rho\sigma}\equiv \frac{q_1}{2}R\left(g^{\mu\rho}g^{\nu\sigma}-g^{\mu\sigma}g^{\nu\rho}\right)
+\frac{q_2}{2}\left(R^{\mu\rho}g^{\nu\sigma}-R^{\mu\sigma}g^{\nu\rho}+R^{\nu\sigma}g^{\mu\rho}-R^{\nu\rho}g^{\mu\sigma}\right)
+q_3R^{\mu\nu\rho\sigma},
\end{equation}
where $q_1$, $q_2$, and $q_3$ are the phenomenological parameters which describe the coupling strength of nonminmal Yang-Mills fields.
Under the Wu-Yang ansatz, the Yang-Mills potential has only one nonvanishing component, $A^{(\varphi)}_{\theta}=-\nu\sin\theta$, where $\nu$ is a parameter related to magnetic charge, and its corresponding field strength component is $F^{(r)}_{\theta\varphi}=-\nu\sin\theta$.
With the assumption of $q \equiv -q_1$, $q_2=4q$, and $q_3=-6q$, a RBH solution can be obtained, where the regularity requires  $q>0$. The corresponding lapse function has the following form,
\begin{equation}
f_3(r)=1+\left(\frac{r^4}{r^4+2Q_m^2q}\right)\left(-\frac{2M}{r}+\frac{Q_m^2}{r^2}+\frac{r^2}{\ell_{\rm AdS}^2}\right),
\label{lapse NEYM}
\end{equation}
where $Q_m$ defined as $Q_m^2\equiv 4\pi \nu^2$ is the magnetic charge of the RBH. It is easy to check that the lapse function has the following asymptotic forms,
\begin{eqnarray}
f_3(r) &\sim & 1+\frac{r^2}{\ell_{\rm AdS}^2}, \qquad r \rightarrow \infty,
\nonumber\\
f_3(r) & \sim & 1+\frac{r^2}{2q}, \qquad r \rightarrow 0,
\end{eqnarray}
which implies that the nonminimal Einstein-Yang-Mills (NEYM) BH has an AdS core.

It is known that the entropy is usually calculated by the Wald formula in the theories where there are higher curvature terms or the curvatures are coupled to matter fields in a nonminimal formalism. However, in the following context, we show that the Bekenstein-Hawking entropy $H(\lambda,\eta)$ serves as a generalized $c$-function rather than the Wald entropy because its asymptotic behaviors in the UV and IR regions prefer those of a  $c$-function. Therefore, we continue to use Eq.~(\ref{H function}) to study the holographic asymptotic behaviors of the RBHs in both subsections \ref{sec:NEYM} and \ref{sec:EGBNED}.

The Wald formula implies that the entropy of any Killing horizons can be expressed as
\begin{equation}
S=-2\pi\oint_{\Sigma}\left(\frac{\delta {\cal L}}{\delta R_{\mu\nu\rho\sigma}}\epsilon_{\mu\nu}\epsilon_{\rho\sigma}\right){\rm d}V,
\label{Wald formula}
\end{equation}
where ${\cal L}$ stands for the total Lagrangian which corresponds to Eq.~(\ref{action NEYM}) and $\epsilon_{\mu\nu}$  the binormal vector of the codimension-2 Killing horizon. For the causal horizon, the integration surface $\Sigma$ is its $\lambda$-fixed slice, which gives rise to ${\rm d}V={\rm d}V_{\rm ind}$. By using Eq.~(\ref{induced volume}), we rewrite Eq.~(\ref{Wald formula}) as
\begin{equation}
{\rm d}S=-2\pi\left.\left(\frac{\delta {\cal L}}{\delta R_{\mu\nu\rho\sigma}}\epsilon_{\mu\nu}\epsilon_{\rho\sigma}\right)\right|_{\Sigma}{\rm d}V_{\rm ind}
=-2\pi\left.\left(\frac{\delta {\cal L}}{\delta R_{\mu\nu\rho\sigma}}\epsilon_{\mu\nu}\epsilon_{\rho\sigma}\right)\right|_{\Sigma}H(\lambda,\eta){\rm d}V_{\mathbb{H}^{3}},
\label{NED ED 1}
\end{equation}
which gives
\begin{equation}
\frac{{\rm d}S}{dV_{\mathbb{H}^{3}}}=-2\pi\left.\left(\frac{\delta {\cal L}}{\delta R_{\mu\nu\rho\sigma}}\epsilon_{\mu\nu}\epsilon_{\rho\sigma}\right)\right|_{\Sigma}H(\lambda,\eta).
\label{NED ED 2}
\end{equation}
We need the explicit expression of the Wald quantity $\frac{\delta {\cal L}}{\delta R_{\mu\nu\rho\sigma}}\epsilon_{\mu\nu}\epsilon_{\rho\sigma}$ to calculate Eq.~(\ref{NED ED 2}). In the case we are discussing, since the Killing horizon consists of the ingoing null geodesics, the binormal vector has only one independent nonvanishing element, $\epsilon_{uv}$, where $u \equiv t+z^{*}$ and $v \equiv t-z^{*}$ are the outgoing and ingoing Eddington-Finkelstein coordinates, respectively, so the binormal vector can be defined as
\begin{equation}
\epsilon_{\rho\sigma}\equiv \delta_{\rho}^{u}\delta_{\sigma}^{v}-\delta_{\sigma}^{u}\delta_{\rho}^{v}.
\label{NED binormal}
\end{equation}
In order to calculate the Wald quantity related to the binormal vector Eq.~(\ref{NED binormal}), we transform the spacetime metric Eq.~(\ref{metric ansatz z}) into the form with the Eddington-Finkelstein coordinates,
\begin{equation}
{\rm d}s^2= \frac{1}{z^2}\left[-h(z){\rm d}u{\rm d}v+{\rm d}\Omega_2^2\right].
\label{metric ansatz uv}
\end{equation}

 For the NEYM BH, the relevant sector of Lagrangian which contributes to the Wald quantity is $R+4\pi{\cal R}^{\mu\nu\rho\sigma}F^{(a)}_{\mu\nu}F^{(a)}_{\rho\sigma}$, so we compute
\begin{equation}
\frac{\delta R}{\delta R_{\mu\nu\rho\sigma}}\epsilon_{\mu\nu}\epsilon_{\rho\sigma}=g^{\mu\rho}g^{\nu\sigma}\epsilon_{\mu\nu}
\epsilon_{\rho\sigma}=- \frac{2z^4}{[h(z)]^2},
\label{Wald formula Einstein gravity}
\end{equation}
and
\begin{eqnarray}
&&\frac{\delta {\cal R}_{\alpha\beta\lambda\tau}}{\delta R_{\mu\nu\rho\sigma}}\epsilon_{\mu\nu}\epsilon_{\rho\sigma}F^{(a)\alpha\beta}F^{(a)\lambda\tau}\nonumber\\
&=&\frac{q_1}{2}\frac{\delta R}{\delta R_{\mu\nu\rho\sigma}}\left(g_{\alpha\gamma}g_{\beta\tau}-g_{\alpha\tau}g_{\beta\gamma}\right)F^{(a)\alpha\beta}
F^{(a)\lambda\tau}\nonumber\\
&&+\frac{q_2}{2}\left(\frac{\delta R_{\alpha\gamma}}{\delta R_{\mu\nu\rho\sigma}}g_{\beta\tau}-\frac{\delta R_{\alpha\tau}}{\delta R_{\mu\nu\rho\sigma}}g_{\beta\gamma}+\frac{\delta R_{\beta\tau}}{\delta R_{\mu\nu\rho\sigma}}g_{\alpha\gamma}-\frac{\delta R_{\beta\gamma}}{\delta R_{\mu\nu\rho\sigma}}g_{\alpha\tau}\right)F^{(a)\alpha\beta}F^{(a)\gamma\tau}\nonumber\\
&&+ q_3\frac{\delta R_{\alpha\beta\lambda\tau}}{\delta R_{\mu\nu\rho\sigma}}F^{(a)\alpha\beta}F^{(a)\gamma\tau}.
\end{eqnarray}
Combining the above two equations with the metric Eq.~(\ref{metric ansatz uv}), we obtain the Wald quantity,
\begin{equation}
\frac{\delta {\cal L}}{\delta R_{\mu\nu\rho\sigma}}\epsilon_{\mu\nu}\epsilon_{\rho\sigma}=- \frac{2z^4}{[h(z)]^2}-16\pi q \nu^2\frac{z^8}{[h(z)]^2}.
\label{NEYM Wald}
\end{equation}
Next we make two arguments.

The first argument is about the second term on the R.H.S. of Eq.~(\ref{NEYM Wald}), which comes from the contribution of the Wu-Yang monopole ansatz.  This term is divergent with the order of $z^4$ at $z \rightarrow \infty$ (at the center of the RBH) because the limit of the first term, $z^4/[h(z)]^2 \sim 1$, is finite.
As the magnetic monopole is a classical substance, the gauge invariant $F_{mn}^{(a)}F^{(a)mn}$ behaves divergently as $1/r^4$ at $r \rightarrow 0$. The monopole ansatz is actually invalid when $r \rightarrow 0$ because the quantum effect at a short distance is not negligible. 
We introduce a cut-off to avoid the monopole divergence. After introducing the corresponding cut-off to the monopole and redefining the normalization in Eq.~(\ref{NED ED 2}), we find the relation,
\begin{equation}
\frac{{\rm d}S}{dV_{\mathbb{H}^{3}}}\propto H(\lambda,\eta).
\label{proppor}
\end{equation} 
Therefore, the asymptotic form of $\frac{{\rm d}S}{dV_{\mathbb{H}^{3}}}$ can be recovered by  Eq.~(\ref{H function}).

The second argument is an additional explanation about the calculation of causal horizon entropy in subsections~\ref{sec:Hayward} and \ref{sec:Bardeen}. In the two previous subsections, we have adopted Eq.~(\ref{H function}) as the causal horizon entropy of Einstein's gravity. In fact,  this causal horizon entropy should be computed by the following formula,
\begin{equation}
\frac{{\rm d}S}{dV_{\mathbb{H}^{3}}}=-2\pi\left.\left(\frac{\delta {R}}{\delta R_{\mu\nu\rho\sigma}}\epsilon_{\mu\nu}\epsilon_{\rho\sigma}\right)\right|_{\Sigma}H(\lambda,\eta).
\label{NED ED 3}
\end{equation}
In accordance with the Wald quantity Eq.~(\ref{Wald formula Einstein gravity}), the causal horizon entropy should be proportional to the factor of $z^4/[h(z)]^2$ even in the Einstein gravity. Due to this factor, the causal horizon entropy vanishes on the AdS boundary because of $\mathop{\mathrm{lim}}\limits_{z \rightarrow 0}z^4/[h(z)]^2=0$. In order to obtain the correct UV behavior, we introduce a suitable UV cut-off to remove this factor. This is reminiscent of what we normally do in order to obtain the Minkowski metric on the AdS boundary. According to AdS/CFT correspondence, the bulk metric approaches asymptotically the flat Minkowski metric on the $\mathbb{R}^{1,D-2}$ boundary when $z\rightarrow 0$. However, there exists a subtlety in this process. The metric diverges when $z \rightarrow 0$. The simplest way to eschew this divergence is to introduce a UV cut-off, {\em e.g.}, $\varepsilon_{UV}$ to the bulk metric in the Poincar\'{e} patch,
\begin{equation}
{\rm d}s^2 = \frac{1}{z^2}\left[-{\rm d}t^2+{\rm d}z^2+{\rm d}{\bf x}_{D-2}^2\right], \qquad z\rightarrow \varepsilon_{UV},
\label{metric Poincare}
\end{equation}
such that Eq.~(\ref{metric Poincare}) approaches the following metric,
\begin{equation}
{\rm d}s^2 = \frac{1}{\varepsilon_{UV}^2}\left[-{\rm d}t^2+{\rm d}{\bf x}_{D-2}^2\right], \qquad z\rightarrow \varepsilon_{UV},
\end{equation}
which is exactly the Minkowski metric if we redefine the metric normalization. For our case in subsections~\ref{sec:Hayward} and \ref{sec:Bardeen}, by introducing a UV cut-off in $z^4/[h(z)]^2$ at $z \rightarrow 0$ and redefining the normalization of entropy density, we thus recover Eq.~(\ref{H function}) and retain the correct UV behavior of causal horizon entropy.

\begin{figure}[h!]
	\centering
    \includegraphics[width=0.46\textwidth]{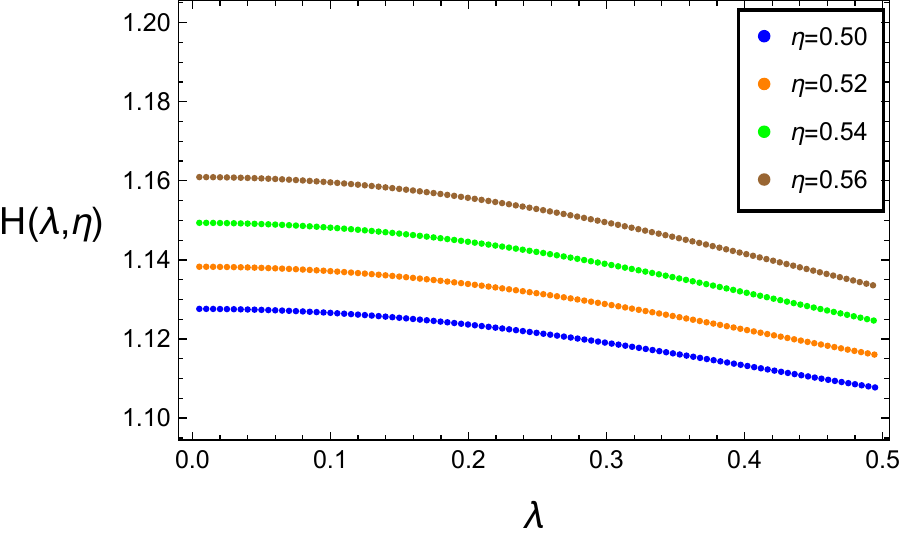}
		\includegraphics[width=0.46\textwidth]{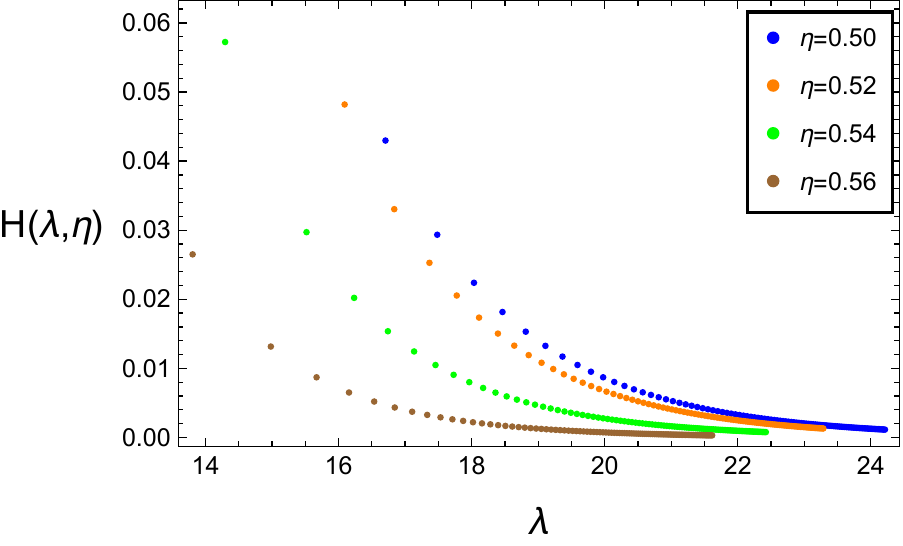}
	\caption{The left panel shows the causal horizon entropy for $z\ll 1$ (UV behavior), while the right panel for $z\gg 1$ (IR behavior). The horizontal axes of both panels represent the affine parameter $\lambda$. We have set the parameters, $M=1$, $q=0.1$, $Q_m=0.1$, and $\ell_{\rm AdS}=1$, for the convenience of numerical calculations and plotting.}
\label{fig:NEYM}
\end{figure}

Finally, for the NEYM BH we plot the UV and IR asymptotic behaviors in Fig.~\ref{fig:NEYM}, from which we can see that the causal horizon entropy successfully manifests the fixed points, especially the IR one at the regular center of NEYM BHs.

\subsection{5D regular charged black hole in Gauss-Bonnet gravity coupled with nonlinear electrodynamics}
\label{sec:EGBNED}

The Gauss-Bonnet gravity coupled with NED fields can be described~\cite{EGBNED} by the action,
\begin{equation}
I_4=\frac{1}{16\pi} \int {\rm d}x^5\sqrt{-g}\left(R-2\Lambda+\alpha{\cal L}_{\rm GB}-4{\cal L}_4(F)\right),
\label{action EGBNED}
\end{equation}
where the cosmological constant $\Lambda=-{6}/{\ell_{\rm AdS}^2}$, and $\alpha$ is the Gauss-Bonnet coupling.
${\cal L}_{\rm GB}$ is the Gauss-Bonnet term given by
\begin{equation}
{\cal L}_{\rm GB}=R^2-4R^{\mu\nu}R_{\mu\nu}+R^{\mu\nu\rho\tau}R_{\mu\nu\rho\tau},
\label{Gauss-Bonnet term}
\end{equation}
and ${\cal L}_4(F)$ is the NED Lagrangian,
\begin{equation}
{\cal L}_4(F)=F\left[1+k_0(-F)^{{1}/{3}}+\frac{23}{18}k_0^2 (-F)^{{2}/{3}}+{\cal O}(k_0^3)\right],
\label{EGB NED quantity 3}
\end{equation}
where $k_0$ is the coupling constant of nonlinear electrodynamics and $F$ has the same definition as that in subsection~\ref{sec:Hayward}. Note that ${\cal L}_4(F)$ approaches the Maxwell Lagrangian in the limit of $k_0 \rightarrow 0$.

One RBH solution can be obtained~\cite{EGBNED}  from Eq.~(\ref{action EGBNED}) together with Eqs.~(\ref{Gauss-Bonnet term}) and (\ref{EGB NED quantity 3}),
\begin{equation}
f_4(r)=1+\frac{r^2}{4\alpha}\left(1-\sqrt{1-\frac{8\alpha}{\ell_{\rm AdS}^2}+\frac{8\alpha q^2}{3kr^4}e^{-{k}/{r^2}}}\right),
\label{lapse EGBNED}
\end{equation}
where $q$ is the reduced charge of the RBH, and $k\equiv k_0\left({q^2}/{2}\right)^{1/3}$. It is quite interesting that Eq.~(\ref{lapse EGBNED}) has the same asymptotic form at both the center and spatial infinity,
\begin{equation}
f_4(r) \sim 1+\frac{r^2}{\ell_{\rm eff}^2}, \qquad  \quad r \rightarrow 0 \quad \textrm{and} \quad r \rightarrow \infty,
\end{equation}
where the effective AdS radius is defined as
\begin{equation}
\ell_{\rm eff}^2\equiv \frac{\ell_{\rm AdS}^2}{2}\left(1-\sqrt{1-\frac{8\alpha}{\ell_{\rm AdS}^2}}\right).
\end{equation}

To calculate the causal horizon entropy, we need the explicit expression of $\frac{\delta {\cal L}}{\delta R_{\mu\nu\rho\sigma}}\epsilon_{\mu\nu}\epsilon_{\rho\sigma}$ for this RBH. From Eq.~(\ref{action EGBNED}), we know that the relevant part is $R+\alpha{\cal L}_{GB}$, where $\frac{\delta R}{\delta R_{\mu\nu\rho\sigma}}\epsilon_{\mu\nu}\epsilon_{\rho\sigma}$ has been given in Eq.~(\ref{Wald formula Einstein gravity}), so we only compute
\begin{eqnarray}
\frac{\delta {\cal L}_{GB}}{\delta R_{\mu\nu\rho\sigma}}\epsilon_{\mu\nu}\epsilon_{\rho\sigma}&=& -4R(g^{uv})^2-8\left(R^{uu}g^{vv}+R^{uu}g^{vv}\right)+2\left(R^{uvuv}-R^{vuuv}-R^{uvvu}+R^{vuvu}\right)\nonumber\\
& \sim & {\cal O}\left(R\right),
\nonumber\\
& \propto & \frac{1}{\ell_{\rm eff}^2}, \qquad z \rightarrow \infty.
\label{EGB Wald 2}
\end{eqnarray}
Note that the contribution of Eq.~(\ref{Wald formula Einstein gravity}) is only a scale factor, $z^4/[h(z)]^2 \sim {\cal O}(1)$, and so is that of Eq.~(\ref{EGB Wald 2}), when $z \rightarrow \infty$.
Therefore, we again recover Eq.~(\ref{H function}) in the IR region with a suitable normalization factor.

\begin{figure}[h!]
	\centering
    \includegraphics[width=0.47\textwidth]{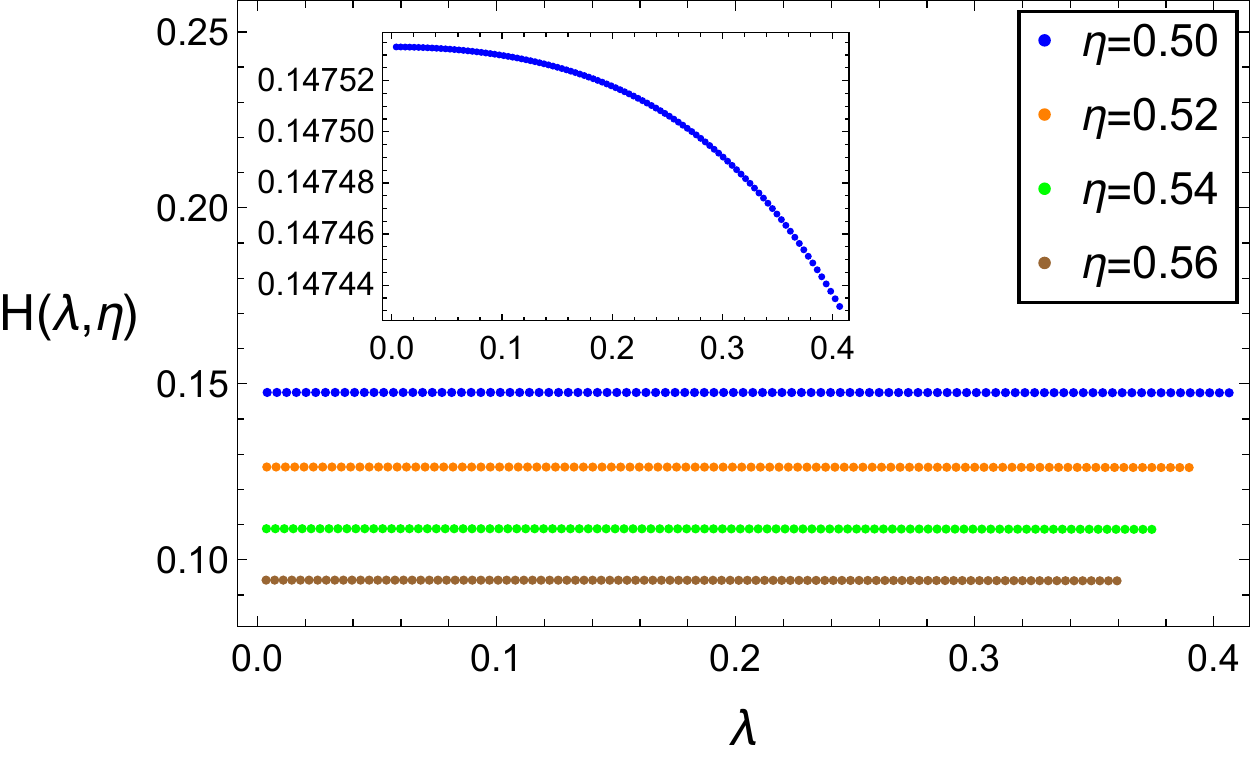}
		\includegraphics[width=0.51\textwidth]{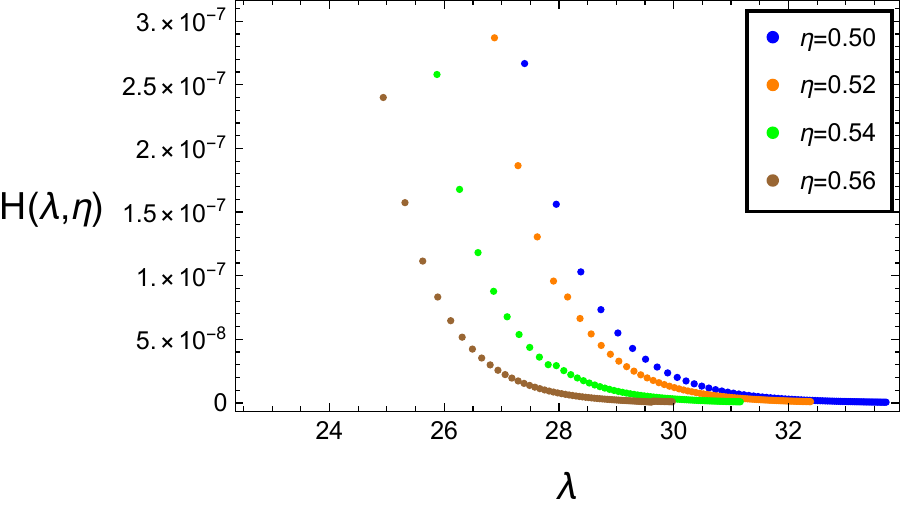}
	\caption{The left panel shows the causal horizon entropy for $z\ll 1$ (UV behavior), while the right panel for $z\gg 1$ (IR behavior). The horizontal axes of both panels represent the affine parameter $\lambda$. We have set the parameters, $\alpha=0.1$, $q=1$, $k=0.1$, and $\ell_{\rm AdS}=0.1$, for the convenience of numerical calculations and plotting.}
\label{fig:EGBNED}
\end{figure}

We plot $H(\lambda,\eta)$ with respect to $\lambda$ for this RBH in Fig.~\ref{fig:EGBNED}. These curves decrease to zero smoothly and are parallel to the horizontal axis, which implies that the causal horizon entropy manifests the IR fixed point at the regular center, and thus it is a valid candidate for the holographic $c$-function of the regular black hole under discussion.

At the end of section~\ref{sec:RBH} we make a summary. Each of the four RBHs has an anti-de Sitter core, therefore its regular center corresponds to an IR fixed point. The causal horizon entropy is a valid candidate of the holographic $c$-function for all the four kinds of black holes. In the next section we shall provide the IR behavior of the Schwarzschild BH (a singular BH) as a comparison.

\section{Schwarzschild black hole}
\label{sec:singular BH}

In this section, we discuss the IR behavior of the Schwarzschild BH which is taken as a sample of singular BHs.

\begin{figure}[h!]
	\centering
    \includegraphics[width=0.49\textwidth]{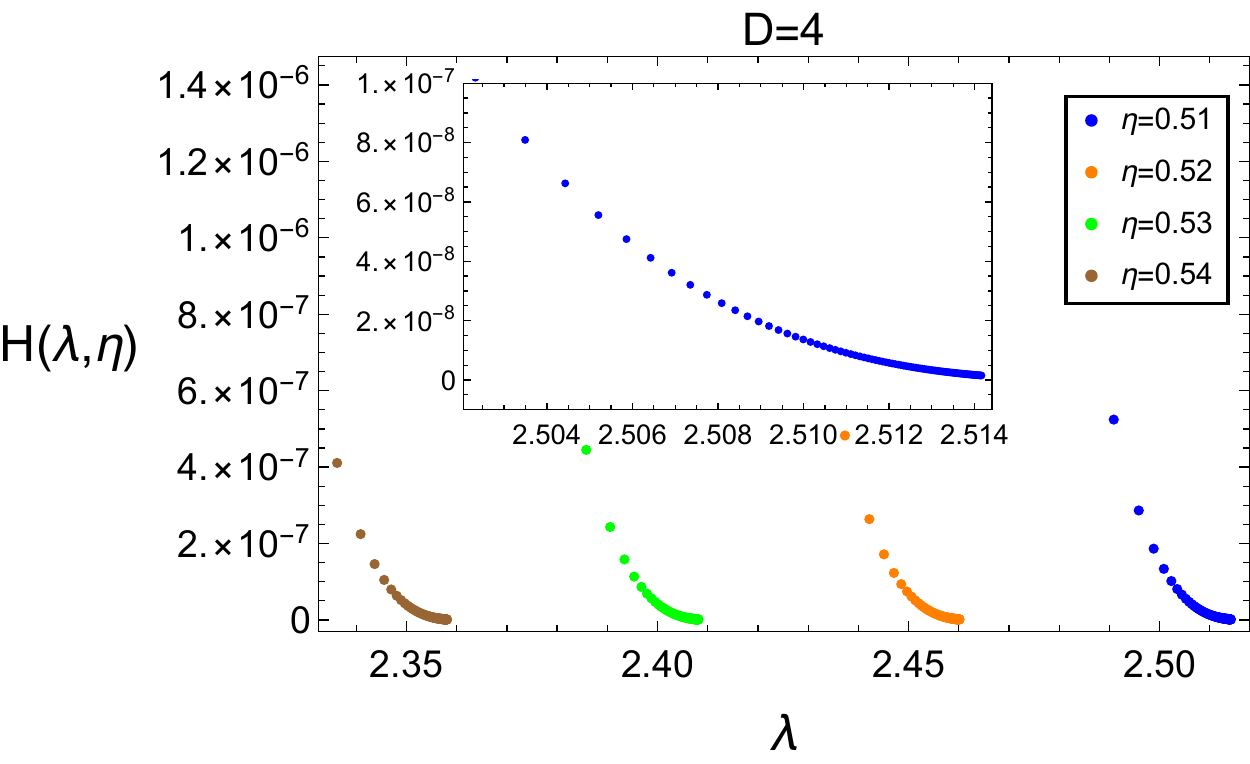}
		\includegraphics[width=0.49\textwidth]{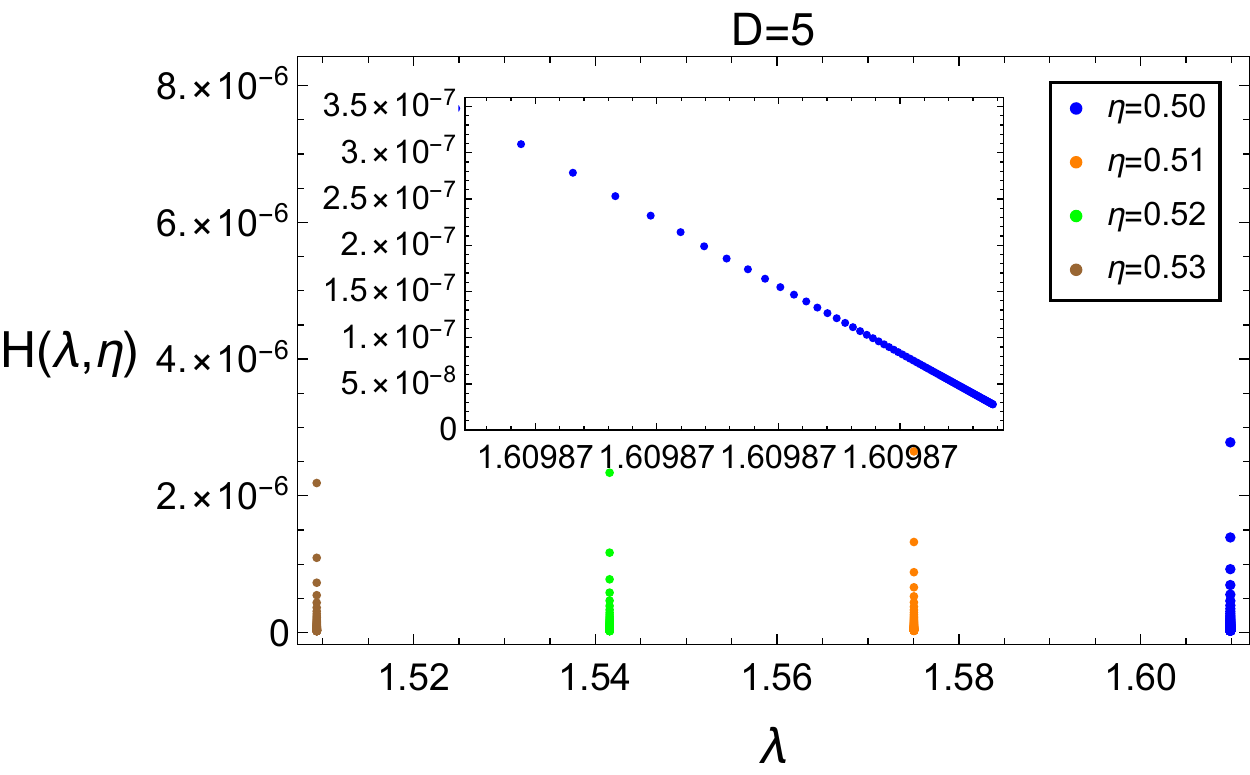}
		\includegraphics[width=0.49\textwidth]{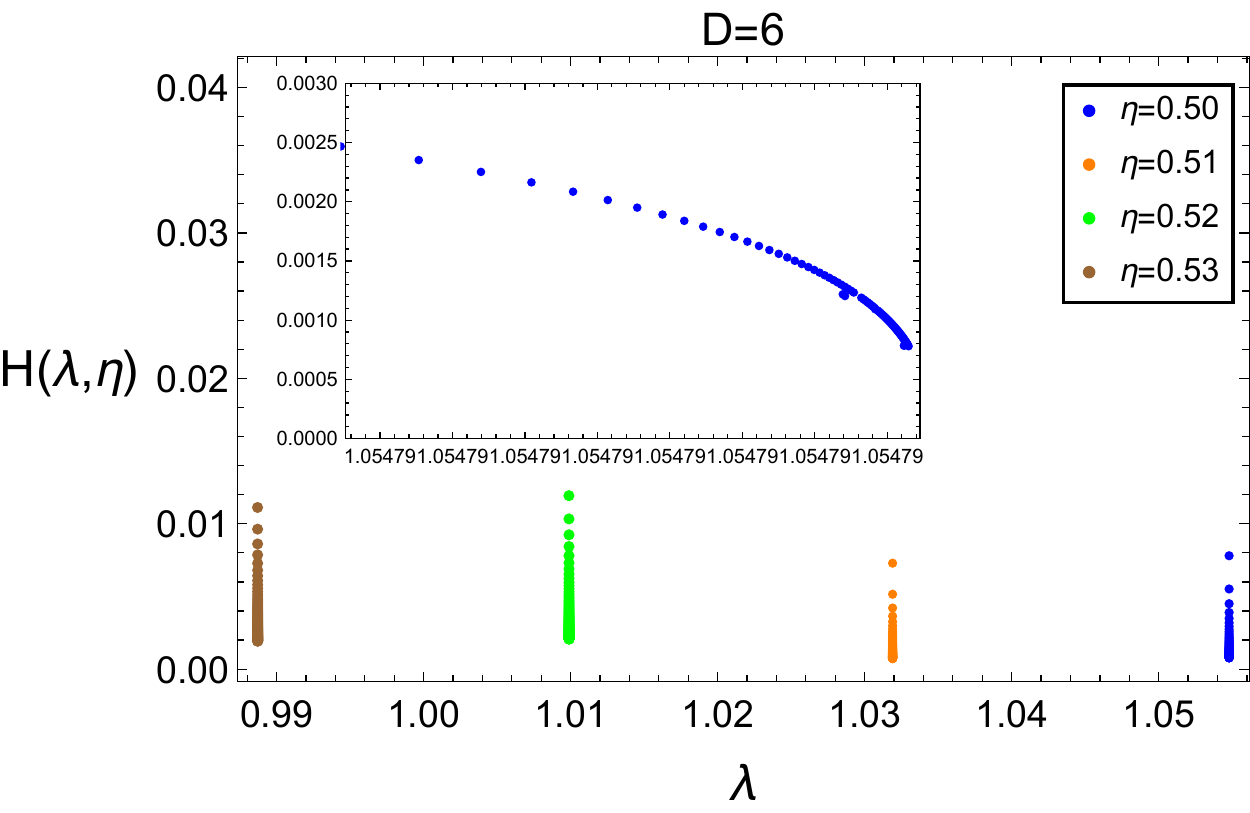}
	\caption{The causal horizon entropy in the Schwarzschild BH for $z \gg 1$ (IR behavior). The horizontal axis represents the affine parameter $\lambda$. We have set the parameters, $m=1$ and $\ell_{\rm AdS}=0.1$, for the convenience of numerical calculations and plotting.}
\label{fig:singular}
\end{figure}

In fact, the stationary IR behavior of a RBH is determined by the lapse function. All of the RBHs we presented in the previous section have an anti-de Sitter core such that
\begin{eqnarray}
f(r) &\sim & 1+ {\cal O}(r^2), \qquad r\rightarrow 0,
\nonumber\\
h(z) &\sim & {\cal O}(z^2), \qquad z\rightarrow \infty,
\label{asymptotic regular}
\end{eqnarray}
which determines that the RBHs have a stationary IR behavior.
In the case of a singular BH, the center is a spacetime singularity, instead of an anti-de Sitter core. The lapse function of a singular BH is divergent at the center such that
\begin{eqnarray}
f(r) &\sim & {\cal O}\left(\frac{1}{r^n}\right), \qquad r\rightarrow 0,
\nonumber\\
h(z) &\sim & {\cal O}(z^{n+2}), \qquad z\rightarrow \infty,
\label{asymptotic singular}
\end{eqnarray}
where $n$ is a positive integer which denotes the leading order of divergence. Due to the asymptotic form Eq.~(\ref{asymptotic singular}), there exist no stationary IR behaviors in the case of singular BHs. For example, the  5D AdS black brane, corresponding to $n=2$, {\em i.e.}, $h(z) \sim {\cal O}(z^4)$,  has no stationary IR behaviors as shown in Ref.~\cite{CH3}. As an additional example, the Schwarzschild BH in four, five, and six dimensions correspond to $n=1, 2, 3$ in Eq.~(\ref{asymptotic singular}), respectively. The lapse function of the Schwarzschild BH is
\begin{eqnarray}
f(r)& =& 1-\frac{a}{r^{D-3}}+\frac{r^2}{\ell_{\rm AdS}^2}, \nonumber\\
h(z) & =& z^2-a z^{D-1}+\frac{1}{\ell_{\rm AdS}^2},
\label{Schwarz lapse}
\end{eqnarray}
where we have adopted a normalized parameter $a$ such that Eq.~(\ref{Schwarz lapse}) has a simple form. We plot the cases of $D=4, 5, 6$ in Fig.~\ref{fig:singular}.

 In the figure the slopes of $H(\lambda,\eta)$ become very steep in the asymptotic IR regions for the cases of $D=5, 6$, and for the case of $D=4$ $H(\lambda,\eta)$ is a concave function and seems to resemble the situation of RBHs. However, by carefully comparing the first graph of Fig.~\ref{fig:singular} with the right graphs of Figs.~\ref{fig:c-Hayward}-\ref{fig:EGBNED}, we find that the curves of the 4D Schwarzschild BH are not exactly stationary because the tangents of them are not parallel to the horizontal axis when $z\to \infty$, which implies nonvanishing derivatives.
 Considering that Fig.~\ref{fig:singular} presents the asymptotic IR region ($z \gg 1$), we conclude that the 4D Schwarzschild BH does not have an IR {\em fixed} point, and that the causal horizon entropy does not have a similar RG flow interpretation as in the cases of regular AdS black holes.

\section{Discussion and conclusion}
\label{sec:concl}

We utilize the causal horizon entropy to study four kinds of RBHs, where the H-function Eq.~(\ref{H function}) serves well as a generalized holographic $c$-function and manifests the asymptotic behaviors of spacetimes. 
The asymptotic behavior of metrics manifests the IR fixed point at the center of regular AdS black holes. Indeed, the causal horizon entropy as a generalized $c$-function displays the properties of the corresponding RG flow. This represents the uniqueness of RBHs. As a comparison, the center of a singular BH is singularity which is excluded from the BH spacetime manifold, and it does not correspond holographically to the IR fixed point. In such a case, the H-function Eq.~(\ref{H function}) loses its RG flow interpretation at the black hole center. Hence, the singularity seems to be an artificial defect of singular BHs, which is caused by an improper construction of gravitational theories. Our analyses indicate that the regular BHs are more likely to be the correct descriptions of BH spacetimes.

It is necessary to emphasize that we use the ``generalized" holographic $c$-function to denominate the causal horizon entropy because of the uniqueness of this $c$-function. There are in total three conditions to construct a $c$-function according to Ref.~\cite{Zamolodchikov}. Here we repeat them and attach an argument behind each of them.

(i) Both ends of a $c$-function correspond to fixed points, which implies that the $c$-function must possess stationary asymptotic behaviors such that its derivative is asymptotically zero, {\em viz.}, $\left.\frac{{\rm d}C(g_i)}{dg_i}\right|_{g_i=g_i^*}=0$, where $C(g_i)$ is a general notation of a $c$-function, $g_i$ is a scale parameter, and $g_i^*$ denotes the RG fixed point.

{\em Our argument}: This property has been verified in section~\ref{sec:RBH}.

(ii) According to the theory of RG flows, any fixed point of the RG flow corresponds to a conformal field theory. The $c$-function is equal to the central charge of the conformal field theory at each fixed point, {\em viz.}, $C(g_i^*)={\cal C}$, where $\cal C$ denotes the central charge.

{\em Our argument}: It is necessary to emphasize that the UV and IR values of our generalized $c$-function do not correspond to the exact values of central charges. These asymptotic values of our $c$-function have a qualitative meaning instead of a quantitative one. Because the center of BHs is a point, which is compactified such that no CFTs exist at this point. However, the IR behavior shown in section~\ref{sec:RBH} is indeed heuristic. In fact, there might be a method to construct an IR CFT. Inspired by the idea of conformal cyclic cosmology~\cite{Penrose,Gurzadyan,Araujo}, we think it is possible to decompactify the BH center $r=0$, analogous to the decompactification of the cosmological singularity. In these literature, the cosmic singularity is decompacified and regarded as the beginning of the new aeon as well as the ending of the old aeon identically. In such a way, a cyclic model of universe is established. Our proposal is to decompactify the center of a RBH and construct an IR CFT there, and further to discuss the quantum correlation between the UV CFT and the IR CFT, and the entanglement between these two CFTs.

(iii) The $c$-function must be monotonic in the RG flow, {\em viz.}, it decreases from UV fixed point to the IR one.

{\em Our argument}: A direct consequence of the monotonicity is that the central charge of the UV theory is larger than that of the IR one. However, this is a necessary but not a sufficient condition of the monotonicity. One can find a function whose UV value is larger than IR value, but this function is not monotonic, such as the causal horizon entropy we have adopted. In fact, the monotonicity is not guaranteed in our models of RBHs. One possible reason is that these models are coupled to the nonlinear gauge field. Due to the nonlinearity, some of the energy conditions may be violated, and thus the monotonicity of causal horizon entropy may be broken~\cite{Jacobson}. Indeed, for the RBHs we focus on above, the numerical method fails on presenting the monotonicity of causal horizon entropy in the region from the UV to IR fixed points because the precision of numerical calculations is only preserved in the two vicinities of UV and IR asymptotic points. Nevertheless, the monotonicity is not our central concern. Recall that the standard $c$-function in four dimensions does not admit monotonicity~\cite{Cardy}, either, but it still manifests the asymptotic behavior of the RG flow --- one fixed point at each end.

In summary, the causal horizon entropy Eq.~(\ref{H function}) is not a holographic $c$-function in the common sense because it is not monotonic. It can be regarded as the measurement of degrees of freedom and quantum correlations~\cite{Myers1}. Despite the absence of monotonicity, our generalized holographic $c$-function successfully manifests the UV and IR behaviors of the RG flow, especially the IR fixed point at the center of RBHs. This implies that the correlation information is not lost at the center of RBHs, in contrast to the situation of singular BHs~\cite{CH3}. We believe that the interpretation of the causal horizon entropy as a generalized holographic $c$-function is a new perspective as for the study of RBHs, which could further enlighten us about the quantum aspects of RBHs.

\section*{Acknowledgements}

The authors would like to thank the anonymous referee for the helpful comments that improve this work greatly. This work was supported in part by the National Natural Science Foundation of China under Grant Nos. 11675081 and 12175108.

\end{document}